# Liquid-phase reinforced Metal matrix (LMM) composite with non-intuitive properties

Varinder Pal[1†], Rakesh Das[1†], Rushikesh Ambekhar[1], Avinash kumar[2], Jitesh Vasavada[3], Banty Kumar[1], Sujoy Kumar Kar[1], Suman Sarkar[4], Mithun Palit[5], Ajit K. Roy[6], Manas Paliwal[7], Krishanu Biswas[8], Sanjit Bhoumik[9], Sushil Mishra[3], Chirodeep Bakli[2], Chandra Shekhar Tiwary[1*]

[1] *Department of Metallurgical and Materials Engineering, Indian Institute of Technology Kharagpur, 721302, West Bengal, India.*

[2] *School of energy science and engineering, Indian Institute of Technology Kharagpur, 721302, West Bengal, India.*

[3] *Department of Mechanical Engineering, Indian Institute of Technology Bombay, 400076, Maharashtra, India.*

[4] *Department of Metallurgical and Materials Engineering, Indian Institute of Technology Jammu, J&K, India.*

[5] *Defence Metallurgical lab, DRDO, Hyderabad, India*

[6] *Materials and Manufacturing Directorate, Air Force Research Laboratory, Wright Patterson AFB, OH 45433-7718, USA.*

[7] *Department of Material Science and Engineering, Indian Institute of Technology Gandhinagar, 382355, Gujarat, India.*

[8] *Department of Metallurgical and Materials Engineering, Indian Institute of Technology Kanpur, UP, India.*

[9] *Bruker, Minneapolis, MN, USA.*

[*]*Email- Chandra.tiwary@metal.iitkgp.ac.in*

# Abstract

Over the ages, efforts have been made to use composite design to reinforce metals and alloys in order to increase their strength and modulus. On the other hand, nature herself improves the strength, ductility, stiffness and toughness of materials by strengthening them with liquids having zero strength/modulus. Here, emulating nature, efforts have been made to develop a new class of tin based alloy/composite with liquid metal reinforcement (LMM). Based on thermodynamic calculations, a composition has been designed such that on melting and casting it forms a solid metal (tin solid solution) and the eutectic mixture remains in liquid form at room temperature. The composite structure named as LMM shows multifold improvement in hardness, strength, ductility, toughness and wear resistance as compared to conventional solder alloys. A Finite Element Method (FEM) based simulation shows strain distribution in the composite which results in the unique behavior. The LMM also shows a negative coefficient of thermal expansion which is further verified using in-situ microscopy and thermodynamic calculations.

# Introduction

Composite materials consist of a matrix and reinforcements (fibrous or particulate etc.), and show superior mechanical properties (high strength to weight and stiffness to weight ratios) as compared to single-phase materials[1,2,3]. The reinforcement's morphology, size, and its interface with matrix governs the mechanical responses (load transfer, strain accommodation, crack propagation etc.) of the composites[4]. Among different classes of composites, metal matrix composites have attracted a lot of attention due to their easy microstructural tunability and high specific strength with reasonable toughness[1,5]. Till date, the mechanical behavior of the metal-based composite is governed by the well-established Eshelby model[6]. As per the model, the modulus/stiffness/strength of the composite is a linear product of mechanical properties (modulus/stiffness/strength) of the individual constituents multiplied by its volume fraction[6]. It implies; that reinforcement with superior mechanical properties (modulus/stiffness/strength etc.) will result in improved mechanical response of the composite. Due to the poor interface[7], the composite (such as carbon nanotube reinforced composite) does not follow the linear improvement law. Hence, various techniques have been developed to engineer the interface for improving the load transfer between the reinforcement and matrix[8,9,10,11].

Going against conventional materials design wisdom, nature chooses a different engineering approach to design composites, which consists of liquid reinforcement in a solid matrix. For example, Barophilic animal's skin, is a natural composite with oil as reinforcement in the compliant solid matrix, which can withstand high pressure and also high impact load[12]. In human, intervertebral discs which are important for the normal functioning of the spine consist of central nucleus pulposus (liquid), a periphery annulus fiber and two vertebral end plates[13] etc.. In all such cases of solid-liquid composites, surface tension, size of liquid reinforcement, and the solid-liquid interface are the key factors affecting stress absorption/strain accommodation[14]. Inspired by nature's design, in the current work we have developed a metal-based composite with liquid phase reinforcement (LMM) and correlated its mechanical response with its structure.

Resistance to corrosion, strength, low melting point and malleability of tin (Sn) alloys (Lead-tin alloys) facilitate their widespread application in the electronics industry as a solder material[15]. The toxicity of lead has led to the development of several lead-free Sn-based alloys[16]. In the current work, we developed a Sn-based alloy with liquid phase reinforcement to mimic nature-based materials design. The Tin-indium-gallium (Sn-In-Ga) ternary alloy system has a eutectic composition ($Ga_{66}In_{20.5}Sn_{9.15}$ wt.% ) with a melting point of 10.7°C[17] and, also a very low vapor pressure as compared to mercury, hence it is used in thermometers[18]. The phase diagram information was obtained through thermodynamic modeling of the Ga-In-Sn system and composition range with primary solid phase and liquid eutectic mixture were known from the calculations. We have developed tin-indium-gallium alloys of the following composition range (Sn- 20-80 wt. %, In- 5-35 wt. % and Ga- 2-80 wt. %). The alloys were synthesized using a simple and easily scalable melting and casting routes. The multifold microstructure characterization of the alloys has been performed at different length scale using optical, electron microscope and X-Ray imaging techniques. The mechanical properties of the alloys have been evaluated using indentation, uniaxial compression and the three-point bend test, and have been compared with tin alloys. The wear properties of the alloys are measured using pin-on-disc and are compared to that of pure tin. To gain insight into the deformation behavior of the unique alloy system, we have performed fracture surface and subsurface deformation studies using scanning electron microscopes. The unique mechanical response is further supported using FEM based simulation. The coefficient of thermal expansion has been measured and compared with solder alloys. An in-situ heating experiment inside SEM has been performed to correlate the coefficient of thermal expansion with microstructure. The negative coefficient is explained with the help of a volumetric calculation using thermodynamic data. The current alloy is the first Sn-based alloy being reported with negative coefficient of thermal expansion which can help in building better interfaced engineered alloys.

## Experimental and simulation details

The thermodynamic calculations for the Sn-In-Ga system were performed with FactSage 7.3 software[19]. Thermodynamic database of the Sn-In-Ga system was prepared by combining the optimized thermodynamic parameters (solution phases and compounds) of the three constituent binary systems: Sn-In, Sn-Ga and In-Sn. The ternary Sn-In-Ga system was optimized by considering the experimental phase diagram for the vertical Ga-17 wt. % Sn – In system determined by Evans and Prince[17]. They also determined the ternary eutectic in the Ga rich side of the phase diagram (Ga-66, In-20.5 Sn-13.5 wt. %, T = 10.7 ±3°C) which was well reproduced in the present study. The thermodynamic parameters for the solution phases (solid and liquid) and the intermetallic compounds for the respective binary systems were taken from the previous studies[20,21,22]. For thermodynamic optimization, the regular solution model was employed to describe the Gibbs energy of liquid and solid solution phases. In order to reproduce the experimental data of Evans and Prince, ternary terms were added to the liquid and Sn solid solution phases.

Alloys with a composition of Sn-14.4 at. % In- X at. % Ga (where X=1.4, 3.4, 5.4, 7.4, 10.4, 13.4, 49.4, 67.4 and 76.4) were synthesized using Sn billet, In block and Ga in liquid form, each with a purity of 99.99%. The alloys were melted using a flame torch for 10 minutes to homogenize the melt as shown in supplementary **fig. S1**. Samples having more than 10.4%Ga were in liquid state at room temperature, so they were not considered for further characterization. Samples for optical microscopy were prepared using conventional metallography techniques and fine polishing using colloidal silica of 0.05μm silica particles. Leica-DM2500M optical microscope was used for imaging of the samples. Imaging at different magnifications was done, and were utilized for calculation of liquid phase fraction. Bruker D8 Advance with Lynx eye detector (Cu-K$_\alpha$) was used for the XRD analysis of the alloys. Melting point of the samples was determined using a differential scanning calorimeter (DSC Q20 V24.2).

The micro-hardness testing was performed using a 4-sided diamond pyramid indenter (UHL VMHT MOT) under different loads (5gf, 10gf and 15gf). The compression testing was performed using a UTM Instron Model 3365 with 0.1mm/min and 0.05mm/min keeping an aspect ratio (height to diameter ratio) of 1.5-2:1. For the three-point bend test, slab samples were prepared as per ASTM E290-14 where C=2r+3t±t/2 (r=radius of Mandrel plunger and t= thickness of sample) is the span length of the sample. Flexural stress (σ) was calculated using the formula "σ= 3FL/2bd$^2$", where F, L, b and d are the load (N), support span (mm), width (mm) of sheet and thickness of the sheet (mm) respectively and the flexural strain (ε) was calculated by using ε= 6Dd/L$^2$, where D, d and L are the maximum deflection of center of beam, thickness of the sheet (mm) and support span (mm) respectively. A steady-state FEM analysis was done to compare the liquid reinforced alloy with solid metal using COMSOL MULTIPHYSICS. Wear rate and coefficient of friction of pure Sn and LLM composite were measured on Make Anton Paar Tribometer,TRB3 (version 8.1.6). Tungsten carbide ball of 6mm diameter was used with a rotating speed of 1cm/s. Scanning electron microscopy (SEM) (Jeol JSM-IT300HR) was used for imaging fractography. Wettability was measured by the contact angles formed by the alloy drops on the printed circuit board (PCB). Thermal mechanical analysis (TMA) was performed on Perkin Elmer Diamond TMA using argon purging at the rate of 50ml/min. The samples for TMA were in disc form with height to diameter ratio within 5:1. The in-situ heating was performed using a heating stage mounted on SEM Pico indenter, PI88 (Bruker, Minneapolis, USA), inside an SEM.

## Results & Discussion

The theoretically calculated ternary phase diagram (Sn-In-Ga), revealing projections of liquidus line is shown in **fig.1 (a).** Different isotherms are shown as a solid line and different phases are marked in the figure. The eutectic composition is shown as a red point in **fig.1 (a)**. The isopleth at constant 14 at. % In composition is shown in **fig.1 (b)**. The presence of the liquid phase along with the solid phase can be observed in the isopleth section and different phases are indicated in the figure. The selected composition for the current study is marked in **fig.1 (a).** The compositions of the alloys studied in the current work consist of L+β, where L and β represents the liquid, and the solid solution of Sn and In respectively. A representative low magnification optical microstructure of solid β phase (body-centered tetragonal) of tin is shown in **fig.1 (c),** revealing single-phase solid microstructures. The microstructure of LMM at different magnifications are shown in **fig. 1(d)-(e)**. We clearly observe the presence of a primary phase of β-tin solid solution surrounded by the ternary liquid eutectic at the boundaries. As we move along the isopleth section with higher concentration of Ga, we observe an increase in liquid fraction as shown in the supplementary **fig. S2**. The calculated liquid fraction of the investigated composites is given in the supplementary **Table 1**. The presence of the phases is further confirmed using the XRD analysis as shown in the supplementary **fig. S3**. The EBSD of the sample shows different orientations of tin in different grains (Inverse pole figure (IPF)) with trapped liquid (black color- without any diffraction) at the grain boundary. We do observe small globules of liquid dispersed in the interior of the grain. The distribution of the trapped liquid is shown using 3D X-ray tomography (**fig.1 (g-h)**). We observe irregular morphology of the liquid on the surface which is found to be connected in the 3D image. The liquid phase mapping using X-ray tomography is shown in **fig. 1(h)**. It clearly reveals that liquid is trapped at the dendritic boundary of the primary solidified Sn-based dendrites. The irregular morphology of the liquid with rough surfaces can be observed in the 3D image.

The average micro-hardness of the LMM composite increases with an increase in the liquid fraction up to 3.18% of the liquid fraction as shown in **fig.2 (a).** It shows the maximum of Vicker's hardness of 35.75±4 VHv. The uniaxial compression test of the LMM composites shows a trend of increase in yield strength with increase in liquid fraction as shown in **fig.2 (b)**. As the fraction of liquid exceeds 8%, there is a fall in the yield strength, but they still has a higher value than that of pure tin (the compressive stress vs strain plot of different LMM composites are provided in the supplementary **fig.S4**).

LMM with 3.4at% Ga shows a maximum yield strength of about 39.38±3 MPa which is higher than that of commercial solder alloys (SAC257, SAC357 and the Sn-37Pb solders). LMM with Ga-5.4% shows an optimum hardness and the yield strength. The flexural stress ($\sigma$) vs strain ($\varepsilon$) curve (**fig.2 (c)**) shows that fracture toughness of LMM composites is higher as compared to that of pure tin. Fractured surfaces of the LMM composites at different length scales (see **fig.2 (d, e, f)**)) shows that the presence of the liquid eutectic phase plays a key role in distributing the stress and avoiding the fracture through the grains (as shown using schematic in **fig.2 (g)**). SEM images of the fractured surfaces shows that the crack is diverted across the eutectic liquid phase (supplementary **fig. S5 and fig. S6**). The liquid tries to restrict/blunt the tip of the crack and hence reduces its propagation. In order to gain insight into the deformation of LMM, a simplified 2-D dimensionless model of 1x1 unit was used for simulation as shown in **fig.2 (h)**. The Sn metal is modelled as a solid square whereas the LMM is modelled as a solid square with non-uniform random grooves for incorporating liquid gallium. A Solid mechanics model was used for simulation with the following boundary conditions: - a uniform load is applied on the upper boundary, the right boundary is taken as Roller boundary condition and symmetry boundary conditions are applied to the left and the bottom boundary. A dimensionless load (p) up to 1000 is applied to see the variation in strain in both models. **Fig. 2 (h)** shows the comparison of volumetric strain between the Sn metal and the LMM composite through contour plots for different loads. It is observed that under the same load conditions, strain in the solid part of the LMM composite is less than in the liquid part. Liquid being a viscous material and trapped in the tortuous cavities gets sheared as the domain is subjected to stress. As opposed to solid, the liquid part does not sustain shear in the modified apertures in the solid part and hence ends up showing greater values of strain for the same applied load. The liquid present in the pores of the composite acts as a cushion which absorbs the energy generated due to load by deforming under load and thus making the solid deformation lesser. This makes the composite more rigid as compared to the Sn metal in which the strain is uniformly distributed (see **Fig.2 (h)**).

The penetration depth ($\mu$m) vs time (sec.) plot for pure tin shows the higher debris formation as compared to the LMM composite **fig.3 (a)**. The large difference in the penetration depth value in LMM is due to the liquid which is coming out of the composite after the initial formation of debris. This liquid is inhibiting the loss of material by forming an envelop above the worn out region. Coefficient of friction ($\mu$) of the LMM composite is 0.05 much lower than that of Sn, which assures the lubrication and formation of coated layer on worn out surface of the LMM composite (see **fig.3 (b)**). The lubricating film can be observed in the SEM images of the LMM composite shown in **fig.3 (c)**. Wettability test of the LMM composite with different liquid fractions is shown in **fig.3 (g)**, the composites with 3.18%liquid and 7.13%liquid shows a minimum contact angle ($\delta$) more than that of pure tin. The melting point is measured using the DSC plot as shown in the supplementary **fig.S7**. As can be seen, a strong endothermic peak is noticeable during the heating cycle with the onset temperature around 100°C (see supplementary **fig. S7)**. Coefficient of thermal expansion ($\alpha$) of Sn and LMM composite has been calculated using using the formula, $\alpha = (L_2-L_1)/(L_1*(T_2-T_1))$, where $L_1$ is the length of sample at temperature $T_1$ and $L_2$ at temperature $T_2$. There is always a positive value of $\alpha$ for pure Sn while the LMM composite shows negative value of $\alpha$ as the temperature goes beyond 85°C. The linear expansion of the pure tin and LMM composite is shown in **fig.4 (c).** This negative change in $\alpha$ is due to the dissolution of solid into the liquid eutectic mixture which can be seen in the SEM images of the LMM composite at room temperature (**fig.4 (d)**), 71°C (**fig.4 (e)**) and 151°C (**fig.4 (f)**). The negative coefficient of thermal expansion of the LMM composite is explained with the help of thermodynamic calculations.

Thermodynamic equilibrium calculation showing the variation of phase amount with temperature for Sn-14.4at%In-5.4at%Ga is shown in **fig.4 (a)**. The calculations suggest that liquid phase formation upon heating is manifested as an endothermic peak during the DSC investigation of this alloy. The onset temperature is well predicted by thermodynamic calculations. For the same alloy, the change in the sample volume as a function of temperature is shown in **fig.4 (c)**. The results for pure Sn are also shown for the purpose of comparison. In case of the ternary alloy, the sample volume is almost constant till 100°C and then it decreases upon further heating. It is important to note that both the thermodynamic calculations and DSC signals suggest the formation of liquid phase around 100°C. With further increase in temperature the sample volume decreases, and the liquid phase fraction increases to almost 0.28 at 140°C. The above experimental evidences suggest a strong correlation between the liquid phase formation and decrease in the sample volume. With increase in temperature beyond 100°C, the decrease of the system volume is induced by phase change, specifically by increase in liquid phase fraction.

Using Maxwell's relations, the coefficient for thermal expansion ($\alpha$) for a system is given as:

$$\alpha = (\partial V/\partial T)_P$$

The above relationship suggests that the coefficient for thermal expansion, under constant pressure, is positive for a system- if the molar volume increases with temperature which is generally the case for solids. In the case of the ternary alloy, the coefficient of thermal expansion is negative beyond 100°C as indicated in experimental data showing volumetric variation of this alloy with temperature. The thermodynamic calculations suggest a phase change with significant amount of liquid phase formation beyond 100°C. As indicated in **fig.4 (b)**, the system consists of liquid phase, β-Sn and ϒ-Sn solid solution, and the molar volume of the system is the weighted sum of the partial molar volume of these phases. Considering that the increase in the liquid phase fraction (beyond 100°C) is compensated by the overall decrease in total solid phase (β-Sn and ϒ-Sn) fraction, it is reasonable to assume that any changes in the molar volume of the system with temperature, is dominated by the liquid phase. From the thermodynamic calculations, it is evident that during heating, the liquid phase is enriched in Sn. Thus, the composition changes in the liquid phase lead to overall decrease of system volume with temperature. In other words, Gibbs energy of the liquid phase becomes more negative with increasing temperature and Sn composition, indicating negative deviations from the ideality. Typically, negative deviations in liquid phase are also indicative of the molar volume of the system being lower than its pure constituents. In order to determine the Gibbs energy of the liquid phase formed during heating of the ternary alloy, thermodynamic calculations were performed, and the result is shown in **fig.4 (a)**. As seen, the Gibbs energy of the liquid phase decreases with temperature, indicating the negative slope of $\partial G/\partial T$. Considering, the negative deviations in the liquid phase, a negative slope of $\partial V/\partial T$ is also expected, which will result in negative coefficient of thermal expansion.

The LMM composite shows an endothermic peak at 140.41°C in DSC curve which is far less than of Sn-37Pb (183°C), SAC (218-220 °C), Sn-5Sb-0.3Cu (220 °C) etc. as shown in **fig.3 (f)**. The hardness of the LMM composite is quite higher than that of other soldering alloys[23,24] without compromising with yield strength[23,25,26] as shown in **fig.4 (g)**. The coefficient of the thermal expansion ($\alpha$) of LMM composite beyond the 85°C temperature is negative in nature which is not seen in any of the Sn-based soldering alloys[23] as shown in **fig.4 (h)**. Contact angle ($\delta$) made by the LMM composite (Sn-14.4In-5.4Ga) on the copper substrate is 52.3° which is higher than that of other soldering alloys[25] (see **fig.3 (h)**). There is a combination of superior properties (yield strength, hardness, coefficient of thermal expansion, low melting point etc.) offered by the LMM composite compared to that of other soldering alloys.

## Conclusion:

Utilizing thermodynamic calculation, the current work has optimized a range of composition in Sn-In-Ga system, which can result into solid-liquid composite (LMM). The alloy compositions are melted using easily scalable conventional melting route and characterize using different microscopy techniques at different length scale. Up to a certain composition, the liquid metal trapped or retained at dendritic/grain boundary is uniformly distributed which results in multifold improvement in hardness, strength and toughness. The FEM analysis shows the heterogeneous strain distribution results in such enhancement. The LMM shows multifold improvement in wear resistance and order of decrease in coefficient of friction. The liquid on ear surface is found to be main reason for such behavior. The coefficient of thermal expansion of LMM is negative, which is possibly originates from melting/dissolution of solid dendrites into liquid. The unique LMM shows superior mechanical, wear and coefficient of thermal expansion as compared to commercial state of art solder alloys. The LMM can be used as electrical/electronic contacts of high speed automotive or space applications.

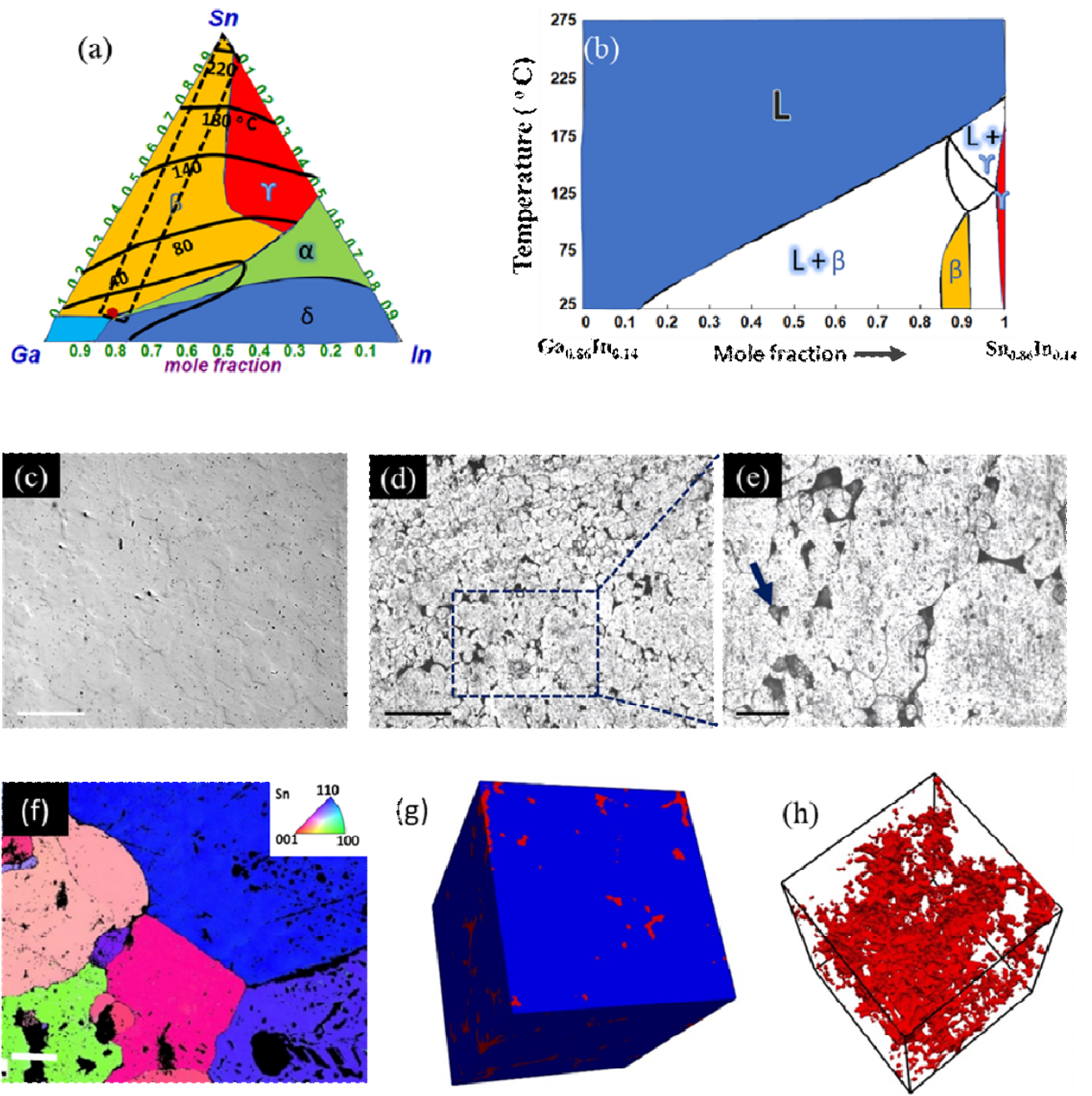

**FIG. 1** (a) The liquidus projections of Sn-In-Ga ternary system, (b) The vertical section in between $Ga_{0.86}In_{0.14}$ and $Sn_{0.86}In_{0.14}$ showing different phases, (c) Optical microstructure of pure tin at 50x magnification, (d) Optical microstructure of $Sn_{78.5}In_{14.4}Ga_{7.4}$ alloy at 50x, (e) Optical microstructure of $Sn_{78.5}In_{14.4}Ga_{7.4}$ alloy at 200x, (f) Inverse Pole figure showing the different grain orientations of $Sn_{80.5}In_{14.4}Ga_{5.4}$ sample at 50µm, (g) Two phase micro CT of $Sn_{80.5}In_{14.4}Ga_{5.4}$, (h) X-Ray Tomography of the $Sn_{80.5}In_{14.4}Ga_{5.4}$ alloy.

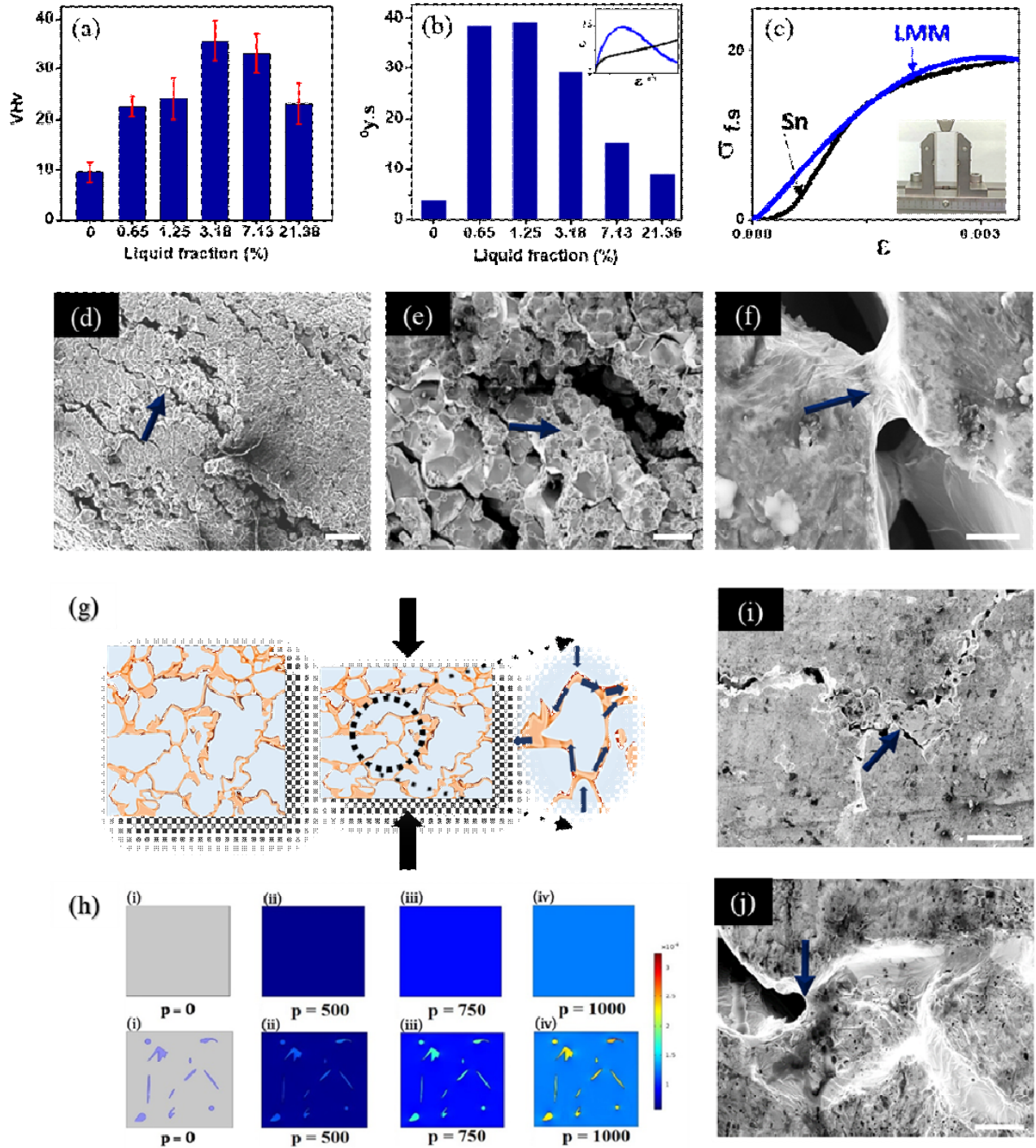

**FIG. 2** (a) Vicker's Micro-hardness (VHv) shows the change in micro-hardness with the increase in fraction of liquid eutectic in the Sn-In-Ga alloys at 5gf load (b) Yield strength($\sigma_{y.s}$) variation with the increase in liquid fraction (c) Flexural stress vs flexural strain curve showing the fracture toughness of the alloys with liquid and without liquid i.e. pure Sn. (d) SEM image of the compressed $Sn_{80.5}In_{14.4}Ga_{5.4}$ sample at 500μm, (e) SEM image of the compressed $Sn_{80.5}In_{14.4}Ga_{5.4}$ sample at 100μm,(f) SEM image of the compressed $Sn_{80.5}In_{14.4}Ga_{5.4}$ sample at 5μm.(g) Schematic of the load transferring through the liquid in compression test of Sn-In-Ga alloy, (h) Volumetric strain contour of solid Sn and LMM composite with variation of dimensionless load (p)=0.500,750,1000. (i) SEM image of the bend surface of Sn80.5In14.4Ga5.4 at 100μm (j) SEM image of the bend surface of Sn80.5In14.4Ga5.4 at 10μm.

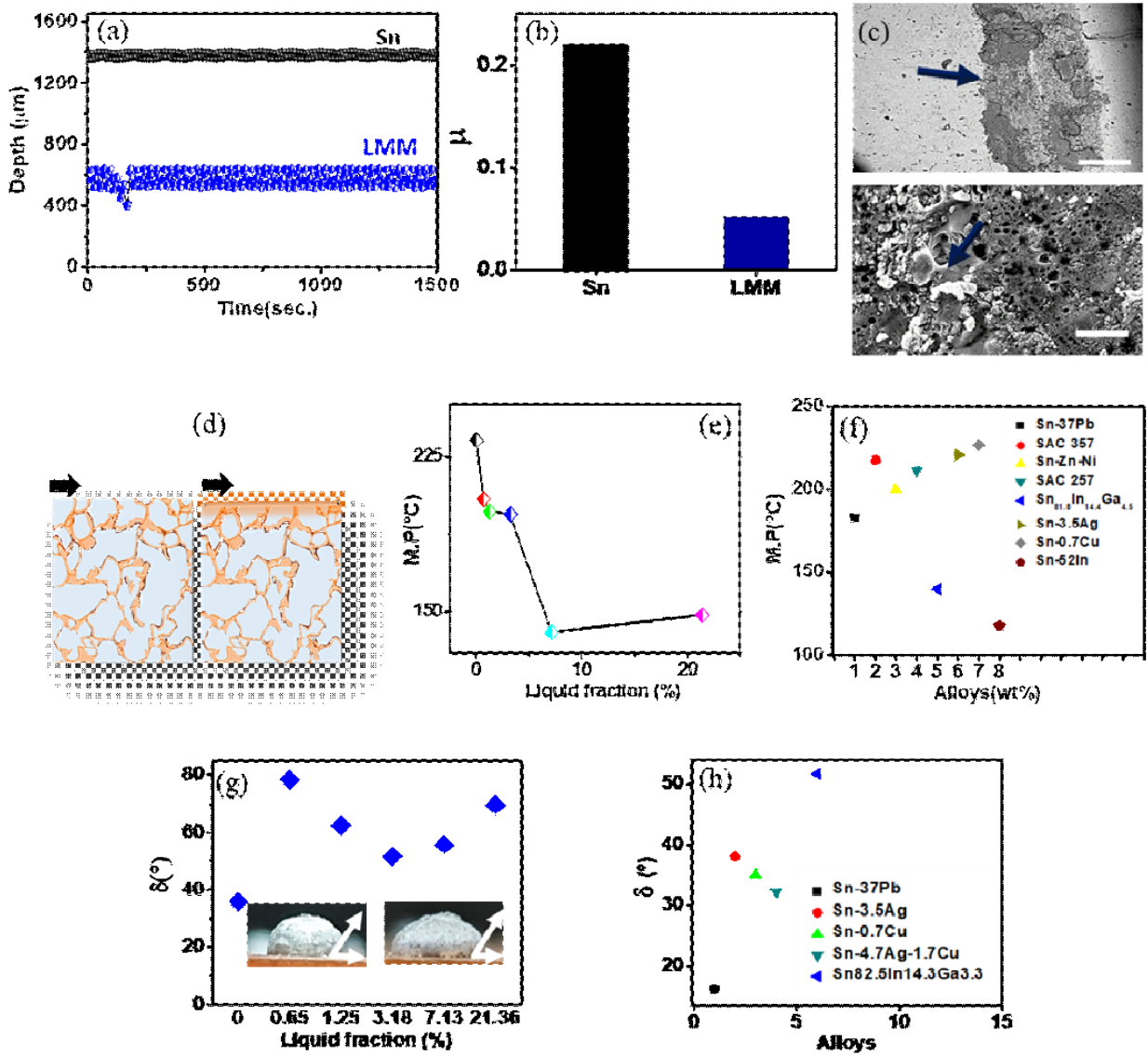

**Fig.3** (a) Penetration depth (μm) of the pure Sn and $Sn_{80.5}In_{14.4}Ga_{5.4}$ (LMM) at a load of 1N with respect to time (seconds), (b) Coefficient of friction (μ) of the pure Sn and $Sn_{80.5}In_{14.4}Ga_{5.4}$ (LMM) on 1N load, (c) SEM image of the worn out surface of LMM composite at 500 μm and 5μm, (d) Schematic of the film formation of the eutectic liquid during the wear testing of the LMM composite , (e) Melting point(°C) vs liquid eutectic fraction (%) of Sn, Sn-1.4a%Ga, Sn-3.4%Ga, Sn-5.4%Ga, Sn-7.4%Ga and Sn-10.4%Ga (at.%) with In-14.4 at.% using differential scanning calorimetric (DSC), (f) Melting point (M.P,°C) of the different solders, (g) Contact angle (δ) made by the Sn-In-Ga alloys with the increase in the liquid fraction in the alloy on PCB substrate, (h) Contact angle (δ) of different soldering alloys on Cu substrate.

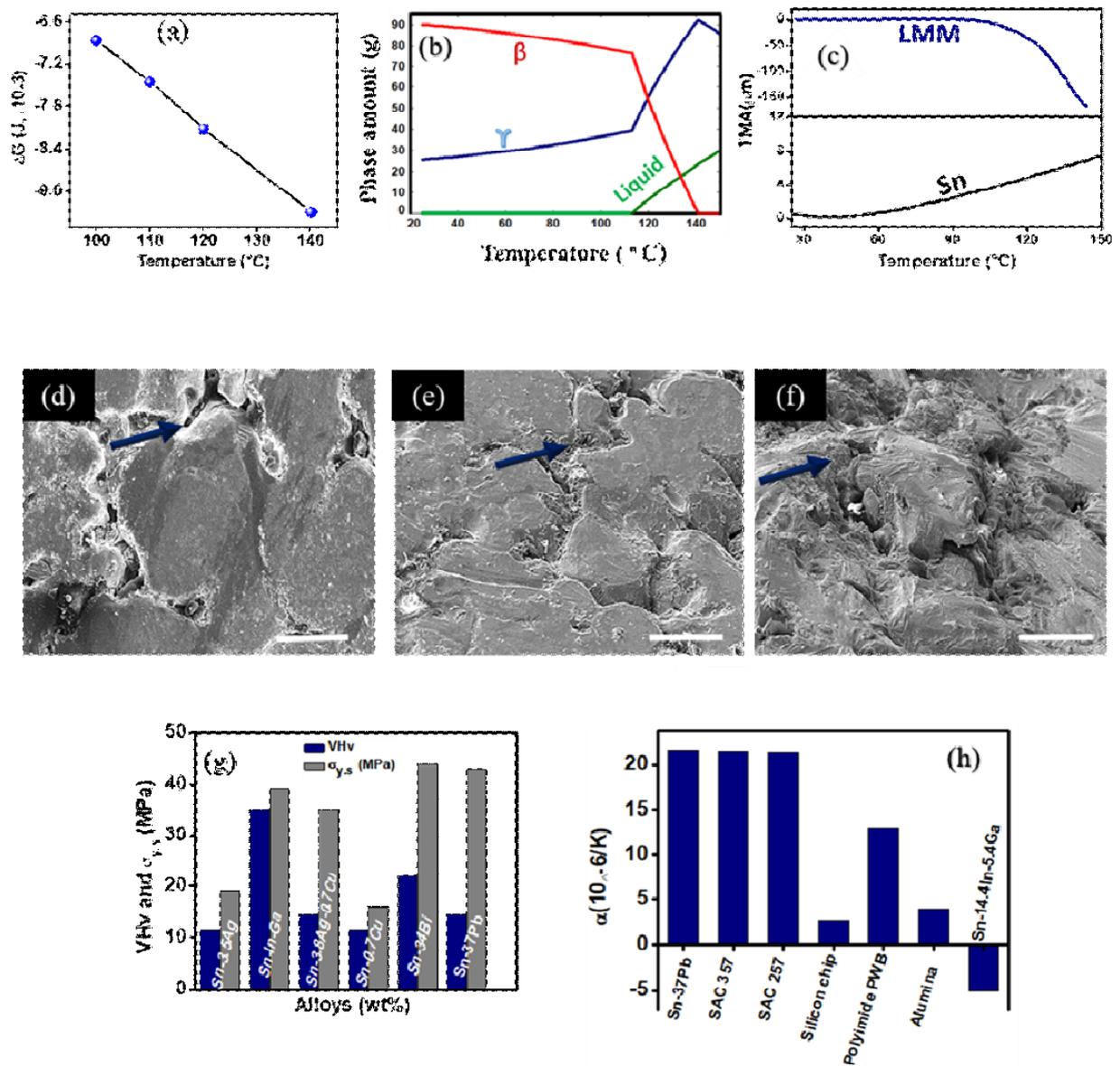

**Fig.4** (a) Change in Gibbs free energy (ΔG, J) plot with respect to temperature from 100°C to 150°C of the $Sn_{80.5}In_{14.4}Ga_{5.4}$ alloy. (b)The phase evolution for the ternary $Sn_{80.5}In_{14.4}Ga_{5.4}$ with temperature (°C), (c) Linear expansion (μm) of Sn and $Sn_{80.5}In_{14.4}Ga_{5.4}$ with increasing temperature (°C), (d) SEM image of the $Sn_{80.5}In_{14.4}Ga_{5.4}$ at ambient temperature at 100μm, (e) SEM image of $Sn_{80.5}In_{14.4}Ga_{5.4}$ at 71°C at 100μm, (f) SEM image of $Sn_{80.5}In_{14.4}Ga_{5.4}$ at 153°C at 100μm. (g) Yield strength ($\sigma_{y.s}$) of the different solder alloys and the Vickers hardness of the different soldering alloys,(h) Coefficient of thermal expansion (α) of various metals and the Sn-In-Ga alloy.